\documentclass[%
reprint,
groupedaddress,
runinaddress,
frontmatterverbose, 
showpacs,showkeys,
bibnotes,amsmath,amssymb,aps,
pre,
]{revtex4-1}

\usepackage{graphicx}
\usepackage{dcolumn}
\usepackage{bm}
\usepackage{textcomp}
\usepackage[mathlines]{lineno}

\begin{document}
%
\title{On the theory of ferroelectric solid solutions }
\author{A.Yu.~Zakharov}\email[E-mail: ]{Anatoly.Zakharov@novsu.ru}
\author{M.I.~Bichurin}\email[E-mail: ]{Mirza.Bichurin@novsu.ru}
\affiliation{Novgorod State University, Veliky Novgorod, 173003, Russia}
\author{S.~Priya}
\affiliation{
 Virginia Tech, USA
}%
%
%
\begin{abstract}
The paper contains an application of lattice model to ferroelectric solid solutions. Short-range parts of interatomic potentials are taken into account by means the lattice structure introduction. Long-range parts are considered in effective field approximation. The equations for the mean values of the components contributions into polarization are derived. The general equation for the Curie temperature dependence on solution composition is obtained. Theoretical calculation of Curie temperature of the system Pb$_{1-x}$Sn$_x$TiO$_3$ fulfilled. 
\end{abstract}
%
\maketitle

\section{Introduction}
Let us consider a ferroelectric solid solution as a lattice with two kinds of dipoles distributed over the sites. Suppose the energy of two dipoles located in sites $\mathbf{R}_i$ and $\mathbf{R}_j$ has the following form~\cite{ZB1,ZB2} \begin{equation}\label{U1}
\begin{array}{r}
{\displaystyle  U_{ij} = - \sum_{\alpha, \gamma}\, Q_{\alpha\gamma}\left(\mathbf{R}_i - \mathbf{R}_j \right)\, \left( \mathbf{D}_{\alpha}\left( \mathbf{R}_i\right) \cdot \mathbf{D}_{\gamma}\left( \mathbf{R}_j\right)\right)  }\\
{\displaystyle \times  n_{\alpha}\left( \mathbf{R}_i\right)\, n_{\gamma}\left( \mathbf{R}_j\right),   }    
\end{array}
\end{equation}
where $n_{\alpha}\left( \mathbf{R}_i\right)$ is a dichotomous random variable with values $1$ and $0$: it is equal to $1$, if site $\mathbf{R}_i$ contains a particle of $\alpha$-th component, and $0$ otherwise.

These variables  $n_{1}\left( \mathbf{R}_i\right)$ and $n_{2}\left( \mathbf{R}_i\right)$ obey the relation
\begin{equation}\label{n12}
    n_{1}\left( \mathbf{R}_i\right) + n_{2}\left( \mathbf{R}_i\right) = 1,
\end{equation}
their mean values are
\begin{equation}\label{<n>}
   \left< n_{1}\left( \mathbf{R}_i\right) \right> = c_1, \qquad \left< n_{2}\left( \mathbf{R}_i\right) \right> = c_2,\qquad c_1 + c_2 = 1
\end{equation}
where $c_1$, $c_2$~ are components concentrations in the solution.

The functions $ Q_{\alpha\gamma}\left(\mathbf{R}\right) $ in lattice models~\cite{Z1,ZaBi} are the long-range parts of interaction potentials with cutting on short distances. A system of dipoles at presence of external electric field $\mathbf{E}$ has the following Hamiltonian
\begin{equation}\label{ham}
\begin{array}{r}
    {\displaystyle  H = - \frac{1}{2}\sum_{i,j}\, \sum_{\alpha, \gamma}\, Q_{\alpha\gamma}\left(\mathbf{R}_i - \mathbf{R}_j \right)\, \left( \mathbf{D}_{\alpha}\left( \mathbf{R}_i\right) \cdot \mathbf{D}_{\gamma}\left( \mathbf{R}_j\right)\right)  }\\
{\displaystyle \times n_{\alpha}\left( \mathbf{R}_i\right)\, n_{\gamma}\left( \mathbf{R}_j\right)     - \sum_{i,\alpha} \left(\mathbf{E}\cdot \mathbf{D}_{\alpha}\left(\mathbf{R}_i \right) \right)\, n_{\alpha}\left(\mathbf{R}_i \right)  }.
\end{array}
\end{equation}
In self-consistent field approximation~\cite{Khach} for the long-range parts of the interactions this Hamiltonian transformed to following form:
\begin{equation}\label{H-eff}
\begin{array}{r}
{\displaystyle   H =  \sum_{\alpha}\, \tilde{H}^{\mathrm{eff}}_{\alpha}\, , }
\end{array}
\end{equation}
where
\begin{equation}\label{H-a-eff}
    \tilde{H}^{\mathrm{eff}}_{\alpha} = -\sum_i \left(\mathbf{E}_{\alpha}^{\mathrm{eff}}\cdot \mathbf{D}_{\alpha}\left(\mathbf{R}_i \right) \right)\, n_{\alpha}\left(\mathbf{R}_i \right),
\end{equation}
\begin{equation}\label{E-eff}
\begin{array}{r}
    {\displaystyle \mathbf{E}_{\alpha}^{\mathrm{eff}} = \mathbf{E}   }\\
{\displaystyle + \frac{1}{2}\, \left< \sum_{j,\gamma}\, Q_{\alpha \gamma}\left(\mathbf{R}_i - \mathbf{R}_j \right)\, \mathbf{D}_{\gamma}\left( \mathbf{R}_j\right) \, n_{\gamma}\left( \mathbf{R}_j\right) \right>.   }
\end{array}
\end{equation}
The last term 
\begin{equation}\label{Ea-eff}
{\displaystyle   \frac{1}{2}\, \left< \sum_{j,\gamma}\, Q_{\alpha \gamma}\left(\mathbf{R}_i - \mathbf{R}_j \right)\, \mathbf{D}_{\gamma}\left( \mathbf{R}_j\right) \, n_{\gamma}\left( \mathbf{R}_j\right) \right>}    
\end{equation}
in this formula is the local effective field at point $\mathbf{R}_i$ due to all the dipoles for dipole of $\alpha$-th kind (in general case the effective fields for the components are not identical).

\section{Polarization evaluation}
\subsection{Effective fields}
To effective fields~(\ref{E-eff}) evaluate it should to find the mean values $\left<n_{\gamma}\left(\mathbf{R}_j \right) \right>$ under the sum over $j, \gamma$ sign in~(\ref{Ea-eff}). In general case these mean values depend on the correlations between components distributions, but in {\em effective field approximation}~\cite{Khach} these mean values are concentrations of the components:
\begin{equation}\label{<n-ij>}
     \left<n_{\gamma}\left(\mathbf{R}_j \right) \right> = c_{\gamma}.
\end{equation}
As a result we have the following expression for effective field~
\begin{equation}\label{E-effc}
    \mathbf{E}_{\alpha}^{\mathrm{eff}} = \mathbf{E} + \frac{1}{2}\,  \sum_{\gamma}\, Q_{\alpha \gamma}^{(0)} \left< \mathbf{D}_{\gamma} \right> c_{\gamma},
\end{equation}
where
\begin{equation}\label{Q0}
    Q_{\alpha \gamma}^{(0)} = \sum_{j}\,Q_{\alpha \gamma}\left(\mathbf{R}_i - \mathbf{R}_j \right) = \sum_{j}\,Q_{\alpha \gamma}\left(\mathbf{R}_j \right)
\end{equation}
(the last relation holds by virtue of translation invariance of crystal).

\subsection{Generating functional and polarization of solution}

Generating functional (i.e. the partition function a system at an external field presence) in effective field approximation has the following form~\cite{Debye,Brown,Froh}:
\begin{equation}\label{Z-sc}
\begin{array}{r}
{\displaystyle   Z = \frac{1}{N_1!\,N_2!} \idotsint \left[\prod_{i,\alpha} \frac{d\Omega_{i,\alpha}}{4\pi} \right]  }\\
{\displaystyle \times  \exp\left[\beta \sum_{i,\alpha}\, \left(\mathbf{E}_{\alpha}^{\mathrm{eff}} \cdot \mathbf{D}_{\alpha}\left(\mathbf{R}_i \right) \right) n_{\alpha} \left( \mathbf{R}_i \right) \right]},    
\end{array}
\end{equation}
where $d\Omega_{i,\alpha}$ is a solid angle infinitesimal element in spherical coordinates $(D_{\alpha}, \theta_i, \varphi_i)$
\begin{equation}\label{dOmega}
d\Omega_{i,\alpha} = \sin\theta_i\, d\theta_i\, d\varphi_i,
\end{equation}
$\beta = 1/T$, $T$ is absolute temperature in energetic units, $N_1$ and $N_2$ are numbers of dipoles.

The multiple integral~(\ref{Z-sc}) is a product of the same type integrals
\begin{equation}\label{int1}
    \int\, \frac{d\Omega}{4\pi}\, \exp\left[\beta {E}^{\mathrm{eff}} D \cos\theta \right] = \frac{\sinh\left( \beta {E}^{\mathrm{eff}} D \right)}{\beta {E}^{\mathrm{eff}} D},
\end{equation}
therefore generating functional in mean field approximation can be evaluated easily
\begin{equation}\label{Z-sc2}
    Z\, = \, \frac{1}{N_1!\,N_2!}\, \left[ \frac{\sinh\left( \beta {E}^{\mathrm{eff}}_1 D_1 \right)}{\beta {E}^{\mathrm{eff}}_1 D_1} \right]^{N_1}\, \left[ \frac{\sinh\left( \beta {E}^{\mathrm{eff}}_2 D_2 \right)}{\beta {E}^{\mathrm{eff}}_2 D_2} \right]^{N_2}.
\end{equation}
Hence we find the system polarization (i.e dipole moment per volume unit)
\begin{equation}\label{P}
\begin{array}{r}
{\displaystyle     P\, = \, \frac{1}{V\beta}\, \frac{\partial \ln Z}{\partial E}\,   }\\
{\displaystyle   = \left\{\frac{N_1}{V} \left[ \coth \left( \beta {E}^{\mathrm{eff}}_1 D_1 \right) - \frac{1}{\left( \beta {E}^{\mathrm{eff}}_1 D_1 \right)} \right]D_1  \right. }\\
{\displaystyle \left. + \frac{N_2}{V} \left[ \coth \left( \beta {E}^{\mathrm{eff}}_2 D_2 \right) - \frac{1}{\left( \beta {E}^{\mathrm{eff}}_2 D_2 \right)} \right]D_2 \right\}  }.
\end{array}
\end{equation}

Thus, for binary solid solution full polarization contains two contributions related two of the components, but these contributions are mutually dependent via effective fields. Therefore the components contributions into polarization are not additive.

\subsection{Mean value of a single particle dipole moment in effective field}

Mean value of a single particle dipole moment in external field $\mathbf{E}$ defined as:
\begin{equation}\label{<D>}
    \left< D \right> = \frac{\int\, e^{-\beta H_1}\, D\, \cos\theta\, d\Omega_1}{\int\, e^{-\beta H_1}\,d\Omega_1},
\end{equation}
and has well known result~\cite{Debye,Brown,Froh}
\begin{equation}\label{<D1>}
   \left< D \right> = D\, L\left(\beta D E \right),
\end{equation}
where
\begin{equation}\label{Lang}
    L\left( z \right) = \left[\coth\left( z \right) - \frac{1}{z} \right]
\end{equation}
is the Langevin function, $H_1$~is the Hamiltonian of single dipole with moment $\mathbf{D}$ in external field $\mathbf{E}$
\begin{equation}\label{H1}
    H_1 = -\left(\mathbf{E} \cdot \mathbf{D} \right).
\end{equation}

For one-component system expression~(\ref{E-eff}) has a simple form
\begin{equation}\label{E-eff1}
   \mathbf{E}^{\mathrm{eff}} = \mathbf{E} + \frac{1}{2}\,  Q^{(0)} \left< \mathbf{D} \right>,
\end{equation}
Substitution $\mathbf{E}^{\mathrm{eff}}$ instead of $\mathbf{E}$ into formula~(\ref{<D1>}) leads to transcendent equation with respect to $\left<D \right>$.
It is well known this equation has nontrivial solution $ \left< D \right> \ne 0$ at $T<T_c$.

\subsection{Polarization of solid solution}

System of equations for mean values  $\left< D_1 \right>$ and $\left< D_2 \right>$ of dipole moments of the solution components follows from relation~(\ref{P}):
\begin{equation}\label{<D1D2>}
\left\{
\begin{array}{r}
    {\displaystyle \left< D_1 \right> =  D_1\, L \left( \beta D_1 \left\{ E + \frac{1}{2} \sum_{\gamma} Q_{1\gamma}^{(0)}\, \left< D_{\gamma} \right>\, c_{\gamma} \right\} \right);   }\\
{\displaystyle  \left< D_2 \right> =  D_2\, L \left( \beta D_2 \left\{ E + \frac{1}{2} \sum_{\gamma} Q_{2\gamma}^{(0)}\, \left< D_{\gamma} \right>\, c_{\gamma} \right\} \right).}
\end{array}
\right.
\end{equation}
Note the interdependence between $\left< D_1 \right>$ and $\left< D_2 \right>$ realizes via non-diagonal elements of matrix  $Q_{\alpha\gamma}^{(0)}$ only.

It is clear that the solution $\left\{\left< D_1 \right>, \left< D_2 \right> \right\}$ of this system of equations depends on temperature, external field, and all the matrix elements $ Q_{\alpha\gamma}^{(0)} $. To solve this system we should know the parameters of the physical system, i.e. matrix elements. This elements should be find using some experimental data.

Solution of system~(\ref{<D1D2>}) with respect to $\left< D_1 \right>$, $\left< D_2 \right>$  permits to find polarization of the system as function of external field $E$
\begin{equation}\label{P2}
    P = \frac{N}{V}\,\left[ c_1\, \left< D_1 \right> + c_2\, \left< D_2 \right>\right]
\end{equation}
and consequently susceptibility $\chi$ of this system:
\begin{equation}\label{chi}
    \chi = \frac{N}{V}\,\left[ c_1\, \frac{\partial \left< D_1 \right>}{\partial E} + c_2\, \frac{\partial \left< D_2 \right>}{\partial E}\right]
\end{equation}

The Langevin function $L(z)$ contains two terms. The first term is algebraic, the second is the transcendent term. Both of them have a singularity at $z=0$. These circumstances complicate search of the solution. Therefore, the Langevin function should be approximated by some more suitable function with correct asymptotic behavior at $z\to 0$ and $z\to\infty$.
In the capacity of such approximation we shall use the following function
\begin{equation}\label{appr}
    L(z) \approx \frac{2}{\pi} \arctan \left(\frac{\pi\, z}{6} \right).
\end{equation}

Note, this approximation is well not only for Langevin function, but also for its derivative
\begin{equation}\label{apprL}
    L^{\,\prime}( z )\approx \frac{12}{36 + \pi^2 z^2}.
\end{equation}

This approximation for Langevin function permits to simplify the system of equations~(\ref{<D1D2>})
\begin{equation}\label{<D-both>}
    \left< D_{\alpha}\right> =  \frac{2\,D_{\alpha}}{\pi} \arctan \left(\frac{\pi \beta D_{\alpha}}{6} \left[  E + \frac{1}{2} \sum_{\gamma} Q_{\alpha\gamma}^{(0)}\, \left< D_{\gamma} \right>\, c_{\gamma} \right]  \right).
\end{equation}

\section{Parameters of the components}

\subsection{Curie temperatures of the components}

Some of the parameters $D_{\alpha}$ and $Q_{\alpha\alpha}^{(0)}$ of solid solution can be find from experimental data of the components. Equation~(\ref{<D-both>}) for a pure component ($c_{\alpha}=1,\, \left. c_{\gamma} \right|_{\gamma \ne\alpha} = 0$) have the following form
\begin{equation}\label{D1}
\begin{array}{r}
    {\displaystyle  \left< D_{\alpha} \right> =  \frac{2\,D_{\alpha}}{\pi} }\\ \\
{\displaystyle \times \arctan \left( \frac{\pi \beta D_{\alpha}}{6}  \left\{ E + \frac{1}{2} Q_{\alpha\alpha}^{(0)}\, \left< D_{\alpha} \right> \right\} \right).  }
\end{array}
\end{equation}

This equation describes connection between external field $E$ and mean value of dipole moment $D_{\alpha}$. Critical temperature can find from condition existence of nontrivial solution (i.e. $\left< D_{\alpha} \right> \ne 0$) in external field vanishing $E=0$. Graphical analysis of the equation~(\ref{<D1>}) leads to following connection  between critical temperature $T_c^{\alpha}$ and model parameters
\begin{equation}\label{T-c}
\frac{\left( D_{\alpha} \right)^2\,  Q_{\alpha\alpha}^{(0)}}{6} = T_c^{\alpha}\,.
\end{equation}

Differentiating both of sides of equation~(\ref{D1}) with respect to $E$
\begin{equation}\label{dDa}
\begin{array}{r}
    {\displaystyle      \frac{\partial \left< D_{\alpha}\right> }{\partial E} = \frac{12 \beta \left( D_{\alpha}\right)^2}{36 + \left(\pi\beta D_{\alpha} \right)^2\, \left[E + \frac{1}{2} Q_{\alpha\alpha}^{(0)} \left< D_{\alpha}\right> \right]^2}  }\\  \\
{\displaystyle  \times  \left(1 + \frac{1}{2} Q_{\alpha\alpha}^{(0)} \frac{\partial \left< D_{\alpha}\right> }{\partial E} \right),  }
\end{array}
\end{equation}
we find derivate
\begin{equation}\label{dDadE}
\begin{array}{r}
{\displaystyle \frac{\partial \left< D_{\alpha}\right> }{\partial E}  } \\
{\displaystyle = \frac{12 \beta \left( D_{\alpha}\right)^2}{ \left[36 + \left(\pi\beta D_{\alpha} \right)^2 \left[E + \frac{1}{2} Q_{\alpha\alpha}^{(0)} \left< D_{\alpha}\right> \right]^2 \right] -  6 \beta Q_{\alpha\alpha}^{(0)} \left( D_{\alpha}\right)^2 }.}     
\end{array}
\end{equation}
Eliminating $Q_{\alpha\alpha}^{(0)}$ from the last equation with account~(\ref{T-c}), we obtain susceptibility of the pure ferroelectrics:
\begin{equation}\label{DE}
\begin{array}{r}
    {\displaystyle  \chi = \frac{N_{\alpha}}{V}\,  \frac{\partial \left< D_{\alpha}\right> }{\partial E}  }\\  \\
{\displaystyle   = \frac{N_{\alpha}}{V} \, \frac{12 \beta \left( D_{\alpha}\right)^2}{ \left\{36\left( 1 - \frac{T_c}{T}\right) + \left(\frac{\pi}{T} \right)^2\, \left[E D + 3 T_c \frac{\left< D_{\alpha} \right>}{D_{\alpha}}  \right]^2 \right\} }.  }
\end{array}
\end{equation}
At $E=0$ the susceptibility have a singularity in vicinity of the critical point $T_c$.

\subsection{Polarization of pure components}

Let us introduce the new variables in equation~(\ref{D1}):
\begin{equation}\label{equi3}
\left\{
\begin{array}{l}
{\displaystyle x_{\alpha} = \frac{\left<D_{\alpha}\right>}{D_{\alpha}}, \quad  \left| x_{\alpha}\right| \le 1; }\\ \\
{\displaystyle \varepsilon = \frac{E D_{\alpha}}{T}; } \\ \\
{\displaystyle \tau = \frac{T}{T_c}. }   
\end{array}
\right.    
\end{equation}
These dimensionless variables have the following physical interpretation: $x_{\alpha}$ is ratio mean value of dipole moment $\left<D_{\alpha}\right>$ to absolute dipole moment $D_{\alpha}$; $\varepsilon$ (dimensionless external field) is ratio the dipole energy $E\,D_{\alpha}$ in external electric field $E$ to temperature $T$; $\tau$ is ratio the temperature $T$ to the Curie temperature $T_c$ defined by~(\ref{T-c}) (thus, $\tau$ is dimensionless temperature).

\begin{figure}
\includegraphics[width=3.0in]{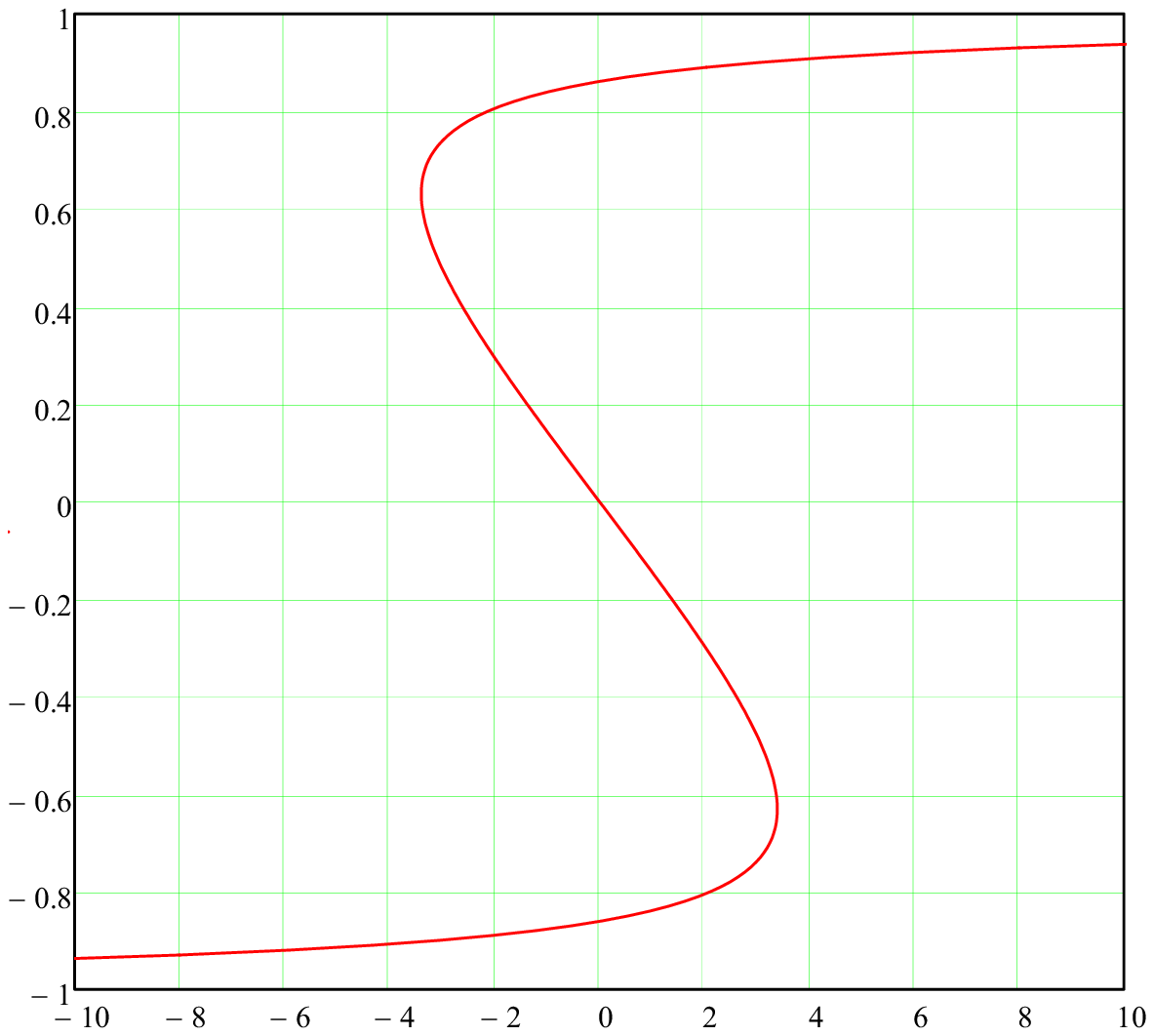}
\caption{Calculated dependence of the dipole moment mean value $x_{\alpha}$ on external field $\varepsilon$ at $\tau=0.3$.}
\label{fig:03}
\end{figure}

\begin{figure}
\includegraphics[width=3.0in]{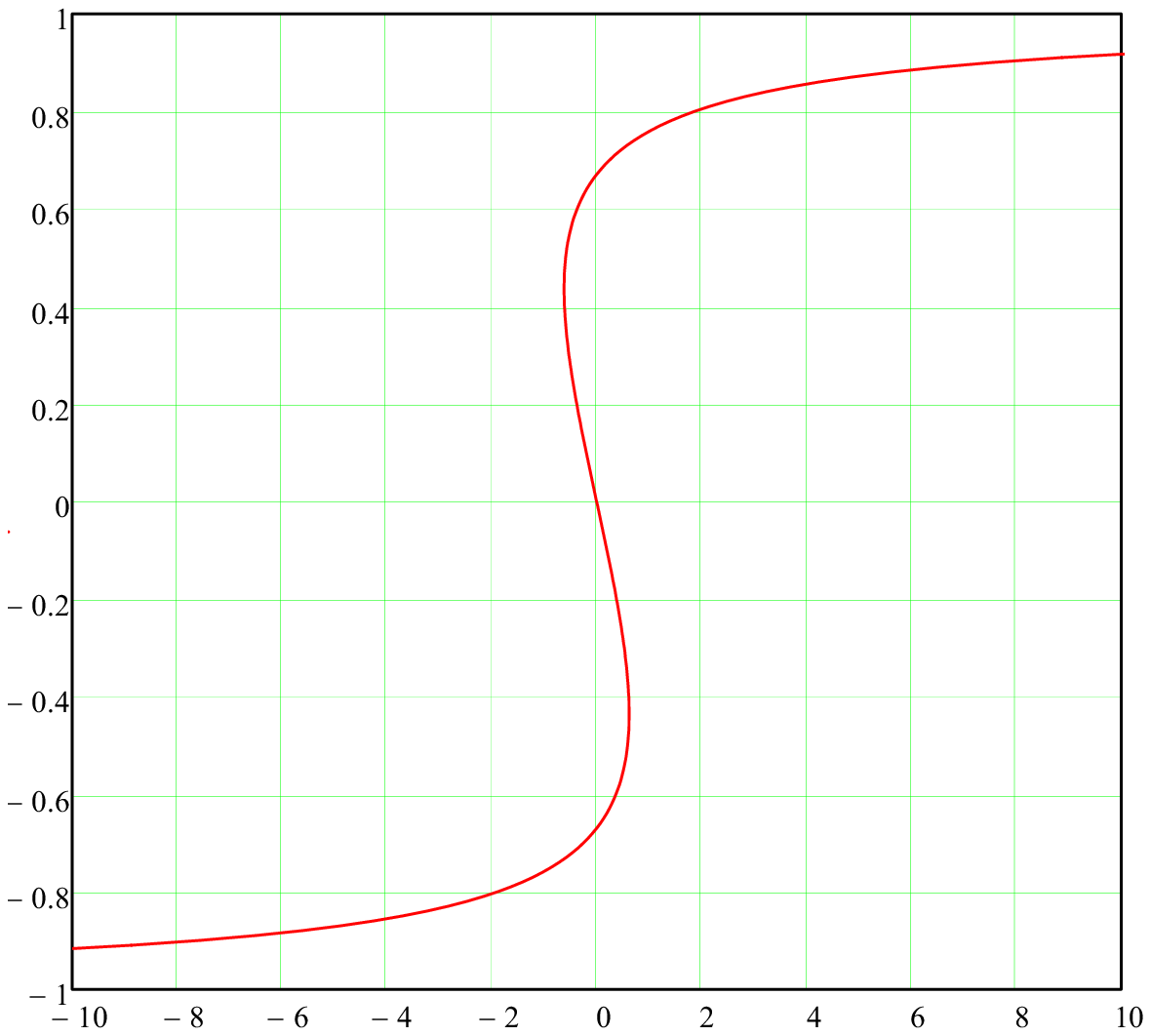}
\caption{Calculated dependence of the dipole moment mean value $x_{\alpha}$ on external field $\varepsilon$ at $\tau=0.6$.}
\label{fig:06}
\end{figure}

\begin{figure}
\includegraphics[width=3.0in]{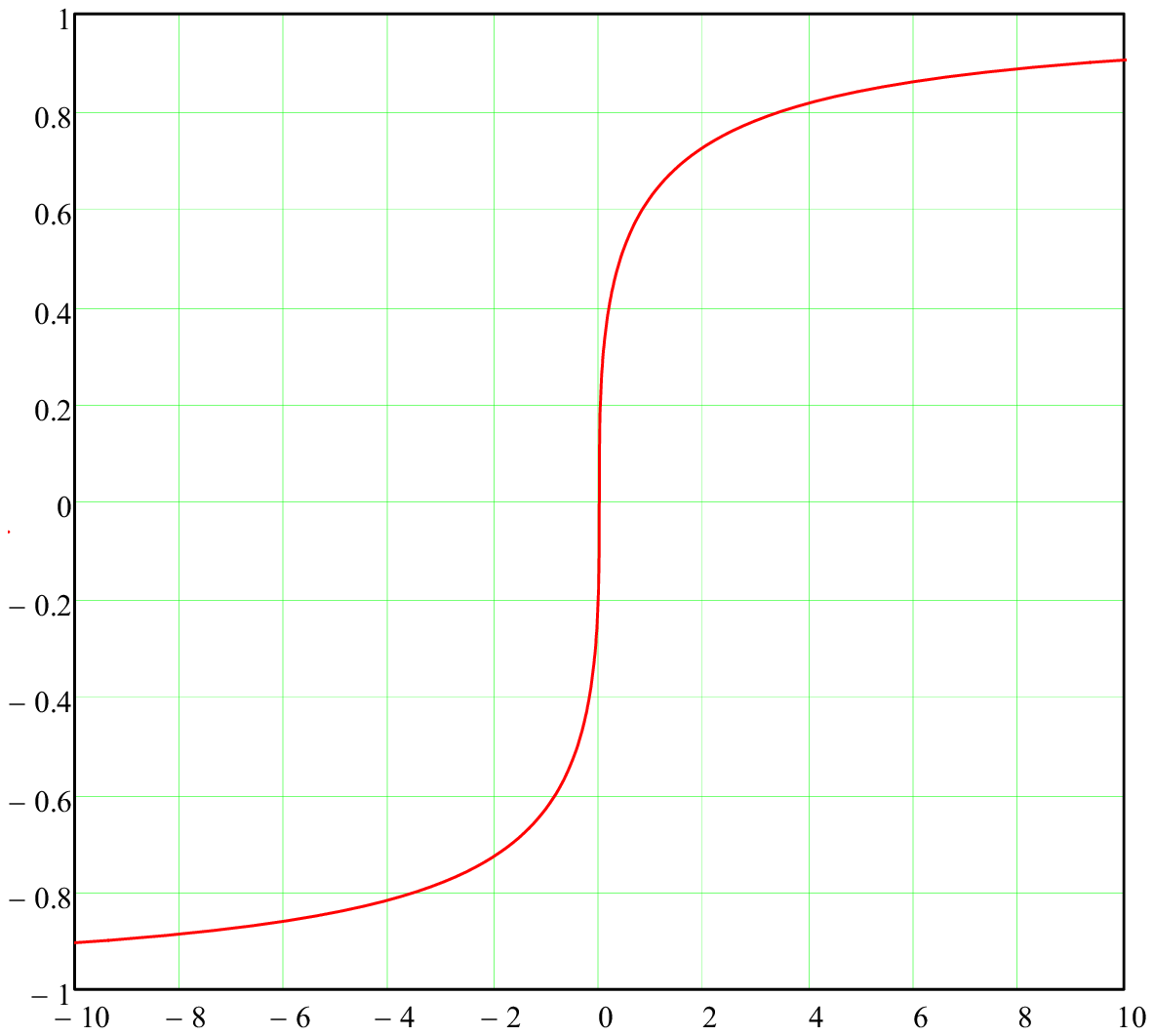}
\caption{Calculated dependence of the dipole moment mean value $x_{\alpha}$ on external field $\varepsilon$ at $\tau=0.9$.}
\label{fig:09}
\end{figure}

\begin{figure}
\includegraphics[width=3.0in]{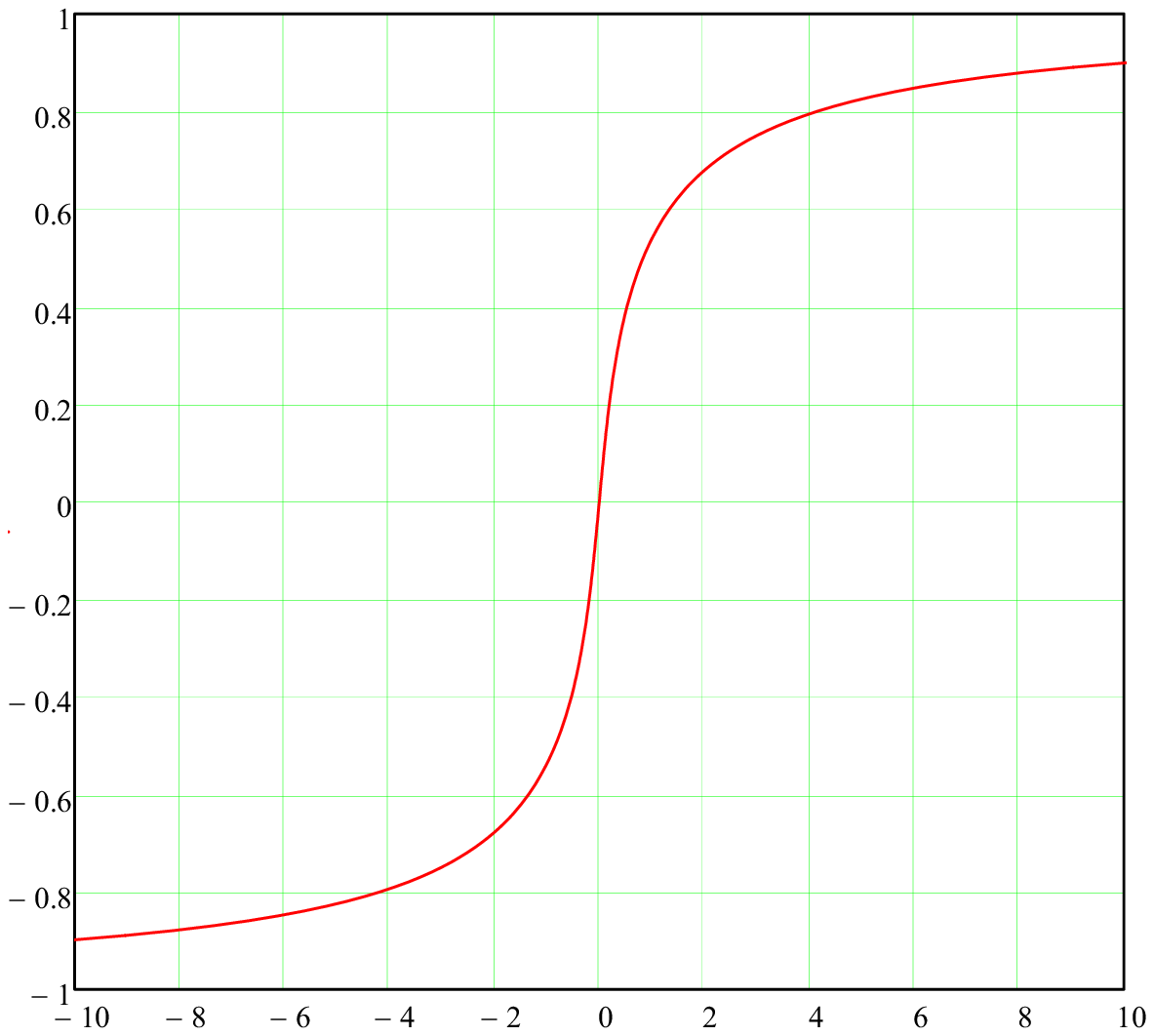}
\caption{Calculated dependence of the dipole moment mean value $x_{\alpha}$ on external field $\varepsilon$ at $\tau=1.2$.}
\label{fig:12}
\end{figure}

\begin{figure}
\includegraphics[width=3.0in]{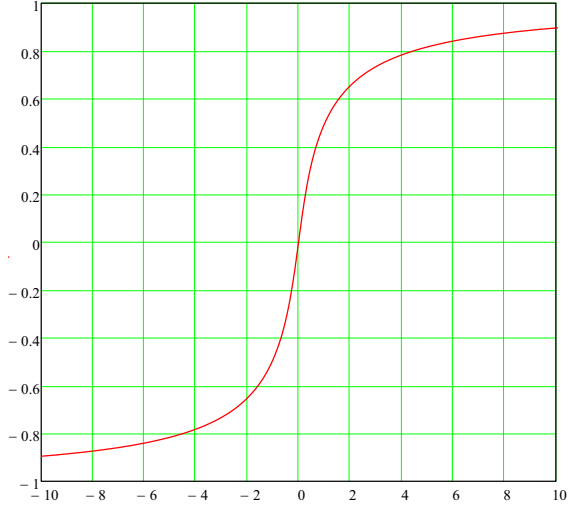}
\caption{Calculated dependence of the dipole moment mean value $x_{\alpha}$ on external field $\varepsilon$ at $\tau=1.5$.}
\label{fig:15}
\end{figure}

Equation~(\ref{D1}) in these variables determines connection between $\varepsilon$, $\tau$, and $x_{\alpha}$:
\begin{equation}\label{x-alpha}
    \varepsilon = \frac{6}{\pi} \tan\left( \frac{\pi}2 \,x_{\alpha} \right) - 3 \frac{x_{\alpha}}{\tau}.
\end{equation}
Connection between dimensionless external field $\varepsilon$ and mean value of dipole moment $x_{\alpha}$ at fixed values of temperature $\tau$ are presented on Figs.~\ref{fig:03}---~\ref{fig:15}.

\begin{figure}
\includegraphics[width=3.0in]{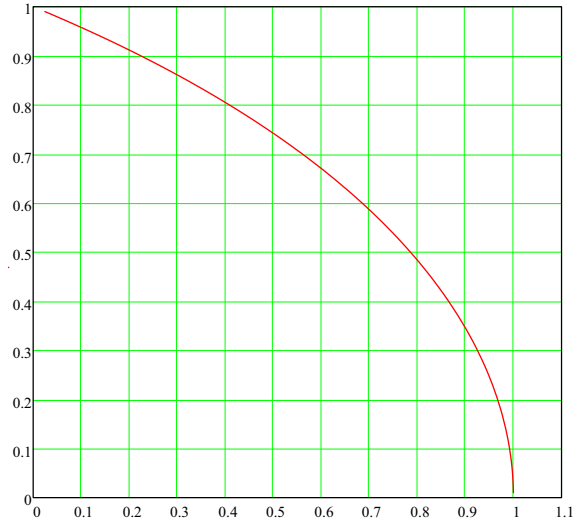}
\caption{Mean value of the dipole moment $x_{\alpha}$ as function on dimensionless temperature $\tau \le 1$.}
\label{x--tau}
\end{figure}

At $T>T_c$ ($\tau>1$) the curves $x_{\alpha}\left(\varepsilon \right)$ pass through the origin. Otherwise (i.e. at $\tau<1)$) there are some spontaneous polarization due to nonzero values of equilibrium mean dipole moments at $\varepsilon=0$. Dependence of dipole moment equilibrium value $x_{\alpha}$ on temperature $\tau$ follows from equation~(\ref{x-alpha}) at $\varepsilon=0$:
\begin{equation}\label{x-tau}
    \tau \tan\left(\frac{\pi x_{\alpha}}{2} \right) - \frac{\pi x_{\alpha}}{2} =0.
\end{equation}
Dependence of the spontaneous value of mean dipole moment $x_{\alpha}$ on temperature $\tau$ presented on Fig.~\ref{x--tau}.
Spontaneous polarization $P$ of pure ferroelectrics expresses via $x_{\alpha}$:
\begin{equation}\label{Px}
    P = \frac{N}{V}D_{\alpha}x_{\alpha},
\end{equation}
where $\frac{N}{V}$ is number of the dipoles $N$ per unity of the system volume $V$.

\section{Susceptibility of solid solutions}

After differentiating both sides of each equation in~(\ref{<D-both>}) over $E$ we obtain a system of linear algebraic equations for $ \frac{\partial \left< D_{\alpha}\right> }{\partial E}$:
\begin{equation}\label{ddE}
\begin{array}{r}
    {\displaystyle     \frac{\partial \left< D_{\alpha}\right> }{\partial E} = \left(1 + \frac{1}{2} \sum_{\gamma} Q_{\alpha\gamma}^{(0)} c_{\gamma} \frac{\partial \left< D_{\gamma}\right> }{\partial E} \right) }\\  \\
{\displaystyle \times \frac{12 \beta \left( D_{\alpha}\right)^2}{36 + \left(\pi\beta D_{\alpha} \right)^2\, \left[E + \frac{1}{2} \sum_{\gamma} Q_{\alpha\gamma}^{(0)} c_{\gamma} \left< D_{\gamma}\right> \right]^2} .   }
\end{array}
\end{equation}
This system of equations has the following short form:
\begin{equation}\label{short}
    \left\{
\begin{array}{l}
    {\displaystyle  \left(1 - A_{1}B_{11} \right)\, \frac{\partial \left<D_1\right> }{\partial E} - A_1 B_{12} \frac{\partial \left<D_2\right> }{\partial E} = A_1;}\\  \\
{\displaystyle -A_2 B_{21} \frac{\partial \left<D_1\right> }{\partial E} + \left(1 - A_{2}B_{22} \right)\, \frac{\partial \left<D_2\right> }{\partial E}  = A_2,}
\end{array}
\right.
\end{equation}
where
\begin{equation}\label{Aa}
    A_{\alpha} = \frac{12 \beta \left( D_{\alpha}\right)^2}{36 + \left(\pi\beta D_{\alpha} \right)^2\, \left[E + \frac{1}{2} \sum_{\gamma} Q_{\alpha\gamma}^{(0)} c_{\gamma} \left< D_{\gamma}\right> \right]^2},
\end{equation}
and
\begin{equation}\label{Bag}
    B_{\alpha\gamma} = \frac{1}{2} Q_{\alpha\gamma}^{(0)} c_{\gamma}.
\end{equation}

Hence we have for $ \frac{\partial \left<D_1\right> }{\partial E} $ and $ \frac{\partial \left<D_2\right> }{\partial E} $:
\begin{equation}\label{D1D2}
\left\{
\begin{array}{l}
{\displaystyle  \frac{\partial \left<D_1\right> }{\partial E}   } \\
{\displaystyle = \frac{A_1 \left(1 - A_2 B_{22} + A_2 B_{12} \right)}{1 - A_1 B_{11} - A_2 B_{22} + A_1 A_2 \left[B_{11} B_{22} - B_{12} B_{21}\right]},  } \\ \\
{\displaystyle  \frac{\partial \left<D_2\right> }{\partial E}   } \\
{\displaystyle = \frac{A_2 \left(1 - A_1 B_{11} + A_1 B_{21} \right)}{1 - A_1 B_{11} - A_2 B_{22} + A_1 A_2 \left[B_{11} B_{22} - B_{12} B_{21}\right]}.  }
\end{array}
\right.
\end{equation}

This system of equations will be used for critical point finding.

\section{Critical points of solid solutions and model parameters}

The solution susceptibility~(\ref{chi}) in the critical point has a singularity due to vanishing denominators in right hand sides of~(\ref{D1D2})
\begin{equation}\label{Tc-equ}
1 - A_1 B_{11} - A_2 B_{22} + A_1 A_2 \left[B_{11} B_{22} - B_{12} B_{21}\right] =0.
\end{equation}
This equation takes place under the conditions
\begin{equation}\label{ED=0}
\left\{
\begin{array}{l}
    {\displaystyle E=0;   }\\
{\displaystyle  \left< D_1 \right>  = \left< D_1 \right>  = 0. }
\end{array}
\right.
\end{equation}
The second of these conditions due to polarization vanishing at $T>T_c$.

Substituting expressions~(\ref{Aa}) and (\ref{Bag}) with conditions~(\ref{ED=0}) account into equation~(\ref{Tc-equ}), we obtain the quadratic equation with respect to critical temperature~$T$:
\begin{equation}\label{T-equ}
\begin{array}{r}
{\displaystyle     T^2 - \frac{\left(D_1 \right)^2 Q_{11}^{(0)} c_1 + \left(D_2 \right)^2 Q_{22}^{(0)} c_2}{6}\, T }\\
{\displaystyle   + \frac{c_1 c_2}{36} \left(D_1 \right)^2 \left(D_2 \right)^2 \left[Q_{11}^{(0)}Q_{22}^{(0)}\, - \, Q_{12}^{(0)}Q_{21}^{(0)} \right] = 0.}    
\end{array}
\end{equation}

The discriminant of this equation in relation to symmetry property of the matrix $Q_{\alpha\gamma}^{(0)}$ is non-negative
\begin{equation}\label{diskr}
\begin{array}{r}
{\displaystyle \tilde{D} =  \left[ \left(D_1 \right)^2 Q_{11}^{(0)} c_1 - \left(D_2 \right)^2 Q_{22}^{(0)} c_2 \right]^2  }\\
{\displaystyle   + 4 c_1 c_2 \left(D_1 \right)^2 \left(D_2 \right)^2 Q_{12}^{(0)} Q_{11}^{(0)} \ge 0,}    
\end{array}
\end{equation}
therefore this equation has two real solutions.

Two variants are possible depending on the quantity Q sign
\begin{equation}\label{Bragg}
    Q = Q_{11}^{(0)}Q_{22}^{(0)}\, - \, Q_{12}^{(0)}Q_{21}^{(0)}.
\end{equation}

\begin{enumerate}
    \item $Q \le 0$. In this case the only of the solutions is positive, and the second solution is negative. The positive solution has a physical sense as Curie temperature, the negative solution has not any physical sense.
    \item $Q > 0$. In this case both of the solutions are positive. A physical interpretation of this case can be clarified after temperature analysis of the equations~(\ref{<D-both>}) solutions. Thus, at condition $Q > 0$ two-component ferroelectric solid solution has two critical points.
\end{enumerate}

Note that the critical points of solutions depends on such combinations dipole moments $D_{\alpha}$ and interatomic potentials $Q_{\alpha\gamma}^{(0)} $
\begin{equation}\label{G-ab}
   G_{\alpha\gamma} = Q_{\alpha\gamma}^{(0)}D_{\alpha}D_{\gamma}.
\end{equation}

Transform the equation (\ref{T-equ}) using these combinations
\begin{equation}\label{Tc}
\begin{array}{r}
{\displaystyle     T^2 - \frac{G_{11} \left(1-x\right) +  G_{22} x}{6}\, T   + \frac{x(1-x)}{36} G = 0,  }
\end{array}
\end{equation}
where 
\begin{equation}\label{G}
    G = G_{11} G_{22}\, - \, G_{12} G_{21},
\end{equation}
$x=c_2$ is second component concentration, $c_1=1-x$.

The left hand side of the equation~(\ref{Tc}) is linear function with respect to  $G_{11}$, $G_{22}$, $G$, therefore the simplest way of the parameters  $G_{11}$, $G_{22}$, $G$ finding from experimental data is the least squares method.

In order to realize this method, let us introduce the function $ f\left(G, G_{11}, G_{22}, T, x\right)$:
\begin{equation}\label{FG}
\begin{array}{r}
{\displaystyle     f\left(G, G_{11}, G_{22}, T, x\right)   }\\  \\
{\displaystyle  =  T^2 - \frac{G_{11} \left(1 - x \right) +  G_{22} x}{6}\, T + \frac{x(1-x)}{36} G  }.
\end{array}
\end{equation}
Then the parameters $G, G_{11}, G_{22}$ can be find by function $F\left(G, G_{11}, G_{22} \right)$ minimization:
\begin{equation}\label{F}
\begin{array}{r}
{\displaystyle      F\left(G, G_{11}, G_{22} \right)  }\\  \\
{\displaystyle  = \sum_{i} f^2\left(G, G_{11}, G_{22}, T_i, x_i\right) = \mathrm{min},  }    
\end{array}
\end{equation}
where $T_i, x_i$ are the experimental points.

Minimization of this function leads to a system of linear algebraic equation with respect to unknown parameters $G_{11}$, $G_{22}$, $G$ and this system can be solved easily.

There is the second way. As a rule, the accuracies of the experimental data for Curie temperatures at points $x=0$ and $x=1$  are higher as at intermediate values of concentrations $x$, hence values of the parameters $G_{11}$, $G_{22}$ should be find by the experimental data of the Curie temperatures of the components using equation~(\ref{T-c}) (with  relation~(\ref{G-ab}) account). The last of the model parameters $G$ can be find using few experimental data of the Curie temperatures $T_c$ of solid solution at intermediate values of the concentrations. 

As an example, let us consider  the solid solution $\mathrm{Pb}_{1-x}\mathrm{Sn}_x\mathrm{TiO}_3$.

\section{System $\mathrm{Pb}_{1-x}\mathrm{Sn}_x\mathrm{TiO}_3$ }

There are the experimental research of the system $\mathrm{Pb}_{1-x}\mathrm{Sn}_x\mathrm{TiO}_3$ in Virginia University. The measurements results presented partially in the table
\begin{center}
\begin{tabular}[c]{||c||c|c|c||} \hline
x & 0.00 & 0.50 & 1.00 \\ \hline
T(x) & 393~K & 623~K & 763~K \\ \hline
\end{tabular}
\end{center}
The first line contains concentration of SnTiO$_3$ in the system, the second line contains the corresponding Curie temperatures in the system.

Unfortunately, the numerical values of the Curie temperatures found from graphics have essential inaccuracies at least order of 10~K. In addition, there is most accurate value of Curie temperature for pure PbTiO$_3$.

Results of the parameters   calculations are: 
\begin{equation}\label{equi2}
\left\{
\begin{array}{l}
{\displaystyle G_{11} = 2358~\mathrm{K}   }\\
 {\displaystyle G_{22}  = 4578~\mathrm{K}; }\\
{\displaystyle  G = -4.043\cdot 10^6~\mathrm{K}^2.}
\end{array}
\right.
\end{equation}

\begin{figure}
\includegraphics[width=3.3in]{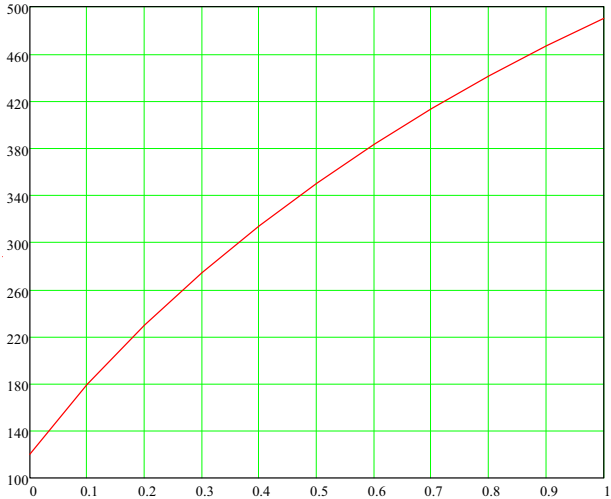}
\caption{Calculated dependence of the Curie temperature (\textcelsius) on the composition $x$ of the system $\mathrm{Pb}_{1-x}\mathrm{Sn}_x\mathrm{TiO}_3$.}
\label{fig:solution}
\end{figure}

Results of these parameters using to Curie temperatures of the solutions are presented on Fig.~\ref{fig:solution}. These results are in close agreement with available experimental data.

\section{Conclusion}
This paper contains the following results.
\begin{enumerate}
    \item Deduction of the closed system of equations with respect to mean values of the elementary dipole moments of the components in ferroelectric solid solutions in effective field approximation for long-range parts of inter-dipole potentials.
    \item The analysis of the connection between polarization, temperature, and external field strength of pure components is fulfilled. These results are presented in analytic and graphical forms.
    \item A way of the parameters model finding from experimental data is presented.
    \item An equation for the Curie temperature of (quasi)-binary ferroelectric solid solutions is obtained.
\end{enumerate}
All of these results allow the obvious generalization on the solutions with arbitrary number of the components. The natural limitation of this approach is assumption on similarity of the components structure.
This approach has some perspectives to composite ferroelectrics and magnetics systems~\cite{Bich1,Bich2}  applications.

\section*{Acknowledgments}
The work was partially supported by the Program of Russian Ministry of Education and Science.

\end{document}